\documentstyle[twocolumn, prl, aps, epsfig]{revtex}
\def\bea{\begin{eqnarray}}
\def\eea{\end{eqnarray}}
\begin{document}

\draft
\title{Order parameter for two-dimensional critical systems with boundaries}

\author{Ivica Re$\check{s}$ and Joseph P. Straley}
\address{Department of Physics and Astronomy,
University of Kentucky,  Lexington, KY 40506-0055}
\maketitle

\begin{abstract}
Conformal transformations can be used to obtain 
the order parameter for two-dimensional systems 
at criticality in
finite geometries with fixed boundary conditions
on a connected boundary.  To the known
examples of this class (such as the disk and
the infinite strip) we contribute the case of 
a rectangle.  We show that the order parameter profile for 
simply connected boundaries can be represented as
a universal function (independent of the criticality
model) raised to the power $ \frac {1}{2} \eta $.
The universal function can be
determined from the Gaussian model or equivalently
a problem in two-dimensional electrostatics.  
We show that fitting the order parameter profile
to the theoretical form gives an accurate route to the
determination of $\eta$.
We perform numerical simulations for the Ising model 
and percolation for comparison with these
analytic predictions, and apply this approach to the
study of the planar rotor model.
\pacs{PACS numbers: $64.60.Cn,05.70.Jk, 64.60.Fr$}
\end{abstract}

\section{INTRODUCTION}
A system undergoing a continuous phase transition is scale invariant
at its critical point.  An
interesting and productive generalization of this idea
for systems also having translational and rotational
invariance 
is to assume conformal invariance\cite{defra,cardy}.  Among other things, this implies
that the magnetization (the one-point correlation function) 
for a two-dimensional critical system transforms
under a conformal transformation $w = w(z)$ in the following way:
\bea
<\phi(w,\bar w)>=(\frac{dw}{dz})^{-h}(\frac{d\bar w}{d\bar z})^{-\bar h}
<\phi(z,\bar z)>.
\label{Eq:eq1}
\eea
Here 
 $< >$ means the 
statistical
average, $z = x + iy$ is a complex-valued coordinate, and it is
assumed that $<\phi(z,\bar z)>$ is nonzero due to a breaking
of the $\phi$ field symmetry at a boundary.  
For a spinless field $\phi$, the scaling exponent
$\Delta$ is related to the conformal dimension h through
the relations
\bea
h = \bar h=\frac{1}{2}\Delta,
\label{Eq:eq2}
\eea
and to the correlation exponent $\eta$ by
\bea
\Delta=\frac{1}{2}\eta
\label{Eq:eq3}.
\eea
Specific values for $\eta$ relevant to what follows are
\cite{tob}
$\eta=\frac{1}{4}$ for the Ising model
and $ \eta=\frac{5}{24}$  for percolation.

In this paper we consider a system at its critical point in a domain, 
with fixed boundary conditions on the boundary.
Examples of the model system we have in mind would be
the magnetization $m(x,y) $ of
the Ising model at its critical temperature defined on the half-plane 
(x,y$>$0), with the spins on the
boundary $y=0$ kept at the value of +1, or a percolating system
at threshold,
where $m(x,y)$ is the probability that the site $(x,y)$
is connected to the boundary $y=0$.  We will describe the
models in this way, even though we will mostly be
considering the continuum versions of the theories,
for which the boundary condition is that
the field is fixed at $\phi \rightarrow \infty $ at 
the boundary.
For the half-plane geometry, the
form of the magnetization is fixed by 
translational, rotational, and scale
invariance 
\bea
<m(x,y)> = Cy^{-\Delta},
\label{Eq:eq4}
\eea
where C is a constant.  This result also describes the behavior
of a lattice model, except when $y$ is as small as a few
lattice spacings; the different behavior at  
short distances can be viewed as the consequence
of corrections to scaling, which in any case are
missing from Eq. (4).
Eq. (1) is relevant to a more general problem, because it relates the
profiles for
different-shaped regions; this has been used\cite{burr} to determine the magnetization profile for several
cases, such as the disk and the infinite strip.  

As will be shown below, conformal invariance
implies that the critical
magnetization profiles for {\em different} models can be
expressed in terms of a single function $m_{1}(\vec r)$:
\bea
m(\vec r) = m_{1}(\vec r)^{\Delta},
\label{Eq:eq5}
\eea
where $m_1 (\vec r)$ is a function that depends on the shape
of the domain but not on any
feature of the model, and describes the
critical behavior of a mythical model with $\Delta = 1 .$
The meaning of this result is that all conformally invariant 
models with the same
exponent $\Delta$ have the same magnetization profile; then it is
enough to look at the magnetization profile for the Gaussian models.
Therefore, Section 2 will begin with a discussion of the Gaussian model and
end with the proof of the assertion (5), and with explicit exhibition of
the function $m_{1}$ for some simple geometries.

Section 3 will present some numerical simulations for the Ising and 
percolation models, verifying that they are related this way and 
showing that fitting the numerical data to the known form of $m_{1}$
gives an accurate way to determine $\Delta$, in which far from worrying
about the implications of boundaries and finite size, we are
{\em making use} of their known effects.

In Section 4 we will demonstrate the usefulness of this technique
in a determination of the temperature
dependence of $\eta$ for the planar spin model.

\section{Order parameter profile in finite geometries according to
the Gaussian model}

The Gaussian model has the action 
\bea
S=\frac {1}{2}g\int|\vec \nabla \theta|^{2}d^2r.
\label{Eq:eq6}
\eea
The order parameter is
\bea
m(\vec r)=Re<e^{i\theta(\vec r)}>,
\label{Eq:eq7}
\eea
where the expectation value can be evaluated from the functional
integral
\bea
m(\vec r)= {{Re \int e^{i\theta(\vec r)}e^{-S}\mathcal D\theta} \over
{\int e^{-S}\mathcal D\theta}}.
\label{Eq:eq8}
\eea
To calculate this integral we expand the field $\theta$ in terms
of a set of functions $\varphi_\lambda$ 
\bea
\theta(\vec r)=\sum_{\lambda }\alpha_\lambda \varphi_\lambda(\vec r),
\label{Eq:eq9}
\eea
and treat the coefficients $\alpha_\lambda$ as independent degrees
of freedom.  The boundary condition on spin ordering becomes the
condition that $\varphi_\lambda(\vec r)$ takes a constant value
(zero, for the moment) for $\vec r$ on 
the boundary.  It proves useful to choose these functions to be the
normalized eigenfunctions $\varphi_\lambda$
of the Laplacian 
\bea
-\nabla^{2} \varphi_\lambda (\vec r)=
\lambda \varphi_\lambda (\vec r),
\label{Eq:eq10}
\eea
because these diagonalize the action; the resulting Gaussian integrals
give
\bea
m(\vec r)=e^{-\frac {1}{2g}\sum_{\lambda }
{\lambda }^{-1} \theta^{2} _\lambda (\vec r)}.
\label{Eq:eq11}
\eea
Interpretation of this result is facilitated by the observation
that the two-point function
\bea
G(\vec R,\vec r)=\sum_{\lambda }\frac{1}{\lambda }\varphi_\lambda(\vec R) 
\varphi_\lambda(\vec r),
\label{Eq:eq12}
\eea
is the Dirichlet Green function: it satisfies
\bea
-\nabla _{\vec R}^{2}G(\vec R,\vec r)=\delta (\vec R,\vec r),
\label{Eq:eq13}
\eea
and gives $
G(\vec R,\vec r)=0$
for $\vec R$ on the boundary.
Thus the quantity appearing in the exponent of (11) appears to 
involve
$lim_{\vec R \to\vec r} G(\vec R,\vec r) $, which 
does not exist.  However, this same difficulty occurs in electrostatics
in the evaluation of the electrostatic energy of a unit point charge at
position $\vec r$ near
a conducting boundary: 
$G(\vec R,\vec r)$ is the potential at point $\vec R$ caused by 
the charge, but to calculate the energy we must remove the self-energy
of the point charge.  In the present context this suggests that we
should replace $G$ by 
\bea
\tilde G(\vec r) = 
lim_{\vec R \to\vec r} [
G(\vec R,\vec r) + 
\frac {1}{2 \pi} ln|\vec R - \vec r| ],
\label{Eq:14}
\eea
and then write
\bea
m(\vec r)=
e^{-\frac {1}{2g} \tilde {G} (\vec r)}.
\label{Eq:eq15}
\eea
This subtraction can be viewed as a modification of the
action (6), in which the fields $\theta$ become
"normal-ordered"\cite{defra}.
What has been subtracted in Eq.(14) is independent of $\vec r$ -- it is 
an (infinite) constant.
The result is that $\tilde G(\vec r)$ is finite throughout
the domain but diverges to $- \infty$ at the boundary,
thus returning the boundary behavior (5) for
$m(\vec r)$.
We see that by solving an
electrostatics problem we can obtain the order parameter
for the Gaussian model defined on the same geometry. 

We now argue that this result describes the
critical magnetization profile for any model that
can be studied by conformal 
field methods.  The argument is as follows:
we can match the order parameter profile 
for the Gaussian model in the case
of the half-plane geometry (Eq. 5) to that of 
any given critical model by choosing  
\bea
\frac {1}{4\pi g}= \Delta,
\label{Eq:eq16}
\eea
since for this case $\tilde G(\vec r) = \frac {1}{2\pi}ln|2y|$ -- 
the part of the Green function due to the "image charge."
But then the conformal transformation rule (1) does not depend on any
property of the ordering field other than the conformal dimension $h$,
or equivalently the exponent $\Delta$.  Thus we can conclude that any
critical model in two dimensions 
will have the same
order parameter profile as its matching Gaussian model, at least 
for all boundary geometries
that are related to the half-plane by a conformal transformation.  
In view of the 
simple way that g enters into Eq. (15),
this result can be restated to say that the order parameter profile
for these cases can be written in the form (5).

Here are some examples:
 
*The disk with
spins fixed at the value of +1 on the boundary:
\bea
<m(r)> = C|A-\frac{r^{2}}{A}|^{-\Delta}.
\label{Eq:eq17}
\eea
Here A is the radius of the disk and r is the distance from the center
of the disk. 
This expression is valid\cite{3D} for both $r < A$ and $r > A$.

*The infinite strip
$\vec r =(x,y)$ with $0 \le y \le W$, with spins fixed at +1 on the 
boundaries $y = 0$ and $y = W$:
\bea
<m(\vec r)> = C(\sin \pi\frac{y}{W})^{-\Delta}.
\label{Eq:eq18}
\eea
The derivation of this expression is discussed in Appendix A.

*The {rectangle $-\frac {L}{2} \le x \le \frac {L}{2}$, $0 \le y \le W$, with spins on all boundaries
fixed at the value +1: 
\bea
<m(w)> = C[{{Im[sn({{2 K} \over {L}} w)]} \over {|cn({{2 K} \over {L}} w)dn({{2 K} \over {L}} w)|}}]^{-\Delta},
\label{Eq:eq19}
\eea
where $w= (x+iy)$ is a complex variable defined on the rectangle, and
sn, cn, dn are the Jacobian elliptic functions\cite{abr}. The aspect
ratio of the rectangle enters through the modulus of the elliptic 
functions: the quarter periods $K$ and $K^\prime$ are related by $K^\prime /K = 2W/L .$ 
The derivation of this result is
given
in appendix A.  Even though elliptic functions are a little
less well known than the functions
that appear in Eqs. (17) and (18), the rectangular geometry is
might be easier to study numerically, and the expression (19) is
easily evaluated.

Common to all the problems mentioned above is that the boundaries
are conformal images of the real axis.  They can also be 
obtained\cite{burr}
by conformal mapping from the half-plane, transforming Eq. (4)
by means of Eq. (1).
 
The order parameter profile for the Gaussian model on an infinite
strip with opposed boundaries (spin +1 on one side, spin -1 on the
other) can be found by an extension of the method.
The Gaussian field $\theta$ is divided into
two parts:
\bea
\theta(\vec r)=\theta_{0}(\vec r)+\theta_{1}(\vec r),
\label{Eq:eq20}
\eea
where $\theta_{1}(\vec r)$ is a fluctuating field that vanishes on the 
boundary while $\theta_{0}(\vec r) = \pi y/W$ 
is the solution to Laplace's equation
with the opposed boundaries.
Partial integration 
then shows 
\bea
&S&=\frac {1}{2}g\int(|\vec \nabla \theta_{0}|^{2}+
|\vec \nabla \theta_{1}|^{2})d^2r,
\label{Eq:eq21}
\eea
so that $\theta_{1}$ is not coupled to $\theta_{0}$ in 
the statistical averaging.  It follows that 
\bea
m_{+-}(\vec r)= < Re
e^{i(\theta_{0}(\vec r)+\theta_{1}(\vec r))}> = 
m_{++}(\vec r)
\cos {\pi y / W },
\label{Eq:eq22}
\eea
where $m_{++} = m_{1}^{\Delta}$ is the profile for 
the case that the
boundary conditions
are the same on opposite boundaries, which we have already considered. 
This result has been previously obtained\cite{burg}
by means of a conformal field
theory argument.
The argument above now implies
that a universal prescription can be given for the magnetization profile
in a region whose boundary is an image of the real axis, with boundary
conditions that change from + to -.   

The Gaussian model is known to give the incorrect two-point
correlation function for the Ising model\cite{cardy}, and so
despite the successes reported above, we should not assume
that the
Gaussian model correctly reproduces the magnetization profile for all
models in {\em all} boundary conditions.  
A counterexample is the case of the Ising model with spins
fixed at +1 on two ends of a cylinder (i.e. a rectangle
with periodic boundary conditions joining a pair of sides).
Since this geometry cannot be related to the half-plane by a conformal
transformation (the boundary is not in any sense connected), 
the chain of logic that allowed us to identify order parameter profiles
for the Gaussian model with those for nontrivial models is broken.  
The Gaussian model solution for this case is
given in Appendix A, with the result
\bea
<m(x,y)>=|\theta_{1}(\frac {2\pi i}{L}y)exp({-\frac {4\pi}{L^{2}\tau}y^{2}})|
^{-\Delta},
\label{Eq:eq23}
\eea
where $\theta_{1}$ is the theta 
function\cite{abr} of
nome $q = exp(- 2 \pi W/L)$ and $\tau=2W/L$.
As will be seen below, simulations do not support this form well,
and indicate that the Ising model magnetization is better 
represented by
\bea
<m(x,y)>=
{{|\theta_{2} ({{\pi i} \over {L}} y )|^{\frac {1}{2}}
+|\theta_{3} ({{\pi i} \over {L}}} y )|^{\frac {1}{2}}
\over
{|\theta_{1} ({{2 \pi i} \over {L}} y)|^{  \frac {1}{8}}}}.
\label{Eq:eq24} 
\eea
In comparing these, please note that $\Delta = \frac {1}{8}$
for the Ising model.
The logic leading to Eq. (24) is explained in Appendix B.

\section{Numerical simulations }

We tested the predictions of conformal field theory by 
simulating the Ising model and the bond percolation
problem on the square lattice.

For the Ising model we used the
Swendsen-Wang Monte Carlo
algorithm \cite {swang}. For each simulation the number of
Monte Carlo steps per site was $10^{6}$. In each 
simulation the intial configuration was all spins at the state of 
+1. The undetermined constant C was used as a fitting parameter.

The percolation threshold for the bond percolation problem
on the square lattice is $p_c=\frac {1}{2} $.  We evaluated
$m( \vec r )$ by averaging $10^{6}$ random bond assignments.

*{\it Disk geometry.}  We cut a disk from a piece of square
lattice with unit lattice constant, defined by the sites
for which $| \vec r | < A $.  Sites outside this disk were
in frozen state: all "spin +" for the Ising model or
"connected to the boundary" for percolation.  We averaged
the magnetization over all directions to obtain $m(r)$.
The results are shown in Figure 1, which is a graph of
$\ln m(r)$ versus $\ln (A - r^{2}/A)$.  The data is expected
to fall on straight lines of slope $- \Delta$, according to
Eq. (17).  Dashed lines having the appropriate slopes have
been drawn in for comparison.  
Since the simulations were done for lattice systems and
the theory is derived for a continuum model, there can
be a discrepancy close to the boundary.  In our simulations,
the boundary is rough at the lattice scale, and in most 
directions is slightly more than $A$ lattice spacings from 
the center.  Therefore we allowed the value used for $A$
in the fitting to be slightly larger than the (integer)
value used to define the disk, chosen so as to eliminate curvature
that would otherwise appear at the left edge of Figure 1
(order close to the disk boundary).  We found 
the optimal
value to be $A = radius + 0.7$ for the Ising model and  
$A = radius + 0.5$ for percolation.
By  fitting the data for the  disks of three different radii
 and finding the average
slope we get $\Delta = 0.1234 \pm0.0005$ for the Ising model, and
$\Delta = 0.1036 \pm0.0002$ for percolation.  The theoretical
prediction from Eq. (17) is $\Delta = 0.125$ for the Ising 
model, and $\Delta \approx 0.10417$ for percolation.  Thus this seems
to be a viable way to determine $\Delta$.

\begin{figure}
\centerline {\epsfxsize=3.in\epsfbox{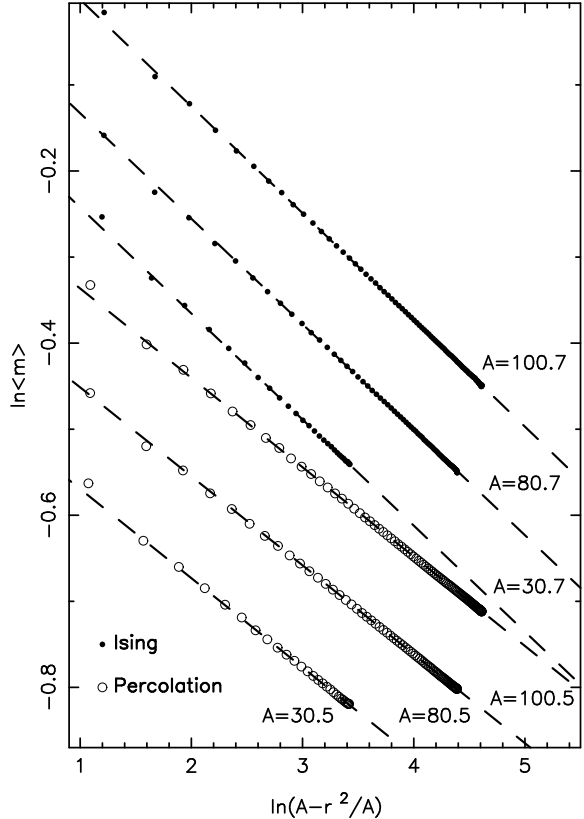}}
\vspace{3mm}
\caption{Linear fits for the Ising model and percolation
on the disk}
\end{figure}

*Ising model on infinite periodic strip with $++$ boundaries. 
In a simulation we can work only with finite strips. 
A rectangle (0$\le$ x$\le$ L,0$\le$ y $\le$ W) 
which is periodic in x direction approximates an infinite strip really well if 
the periodic dimension (in our case L) is greater or equal then the dimension in
the nonperiodic direction (in our example W). The results for the case L = W
are shown in Fig. 2. The spins on the boundary y=0 and y=W were 
fixed at the value of +1.  
\begin{figure}
\centerline {\epsfxsize=\columnwidth\epsfbox{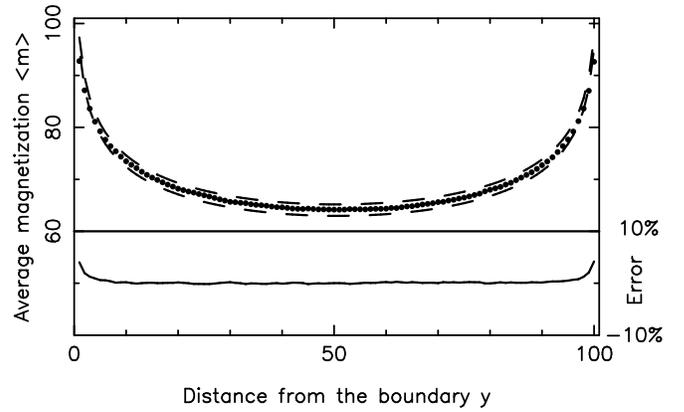}}
\vspace{3mm}
\caption{Average value of magnetization as a function
of the distance from the boundary for L=W=100.}
\end{figure}
The dashed lines are the theoretical curve
\bea
<m(r)> = C(\sin \pi\frac{y}{W})^{-\frac{1}{8}},
\label{Eq:eq25}
\eea
obtained from Eq. (18), displaced by a small constant amount
upwards and downwards, since otherwise the
theory and data would greatly overlap.
The lower part of this figure is a separate graph
of the same information, now presented as the
relative error
$ER=100*({m}_{th}-{m}_{exp})/{m}_{exp}$.
We will use this format in most of our figures.

*Ising model on infinite strip with $+-$ boundaries. 
As above, we approximate the infinite strips with 
rectangles (0$\le$ x$\le$ L,0$\le$ y $\le$ W),L$\ge$W. 
The results for the case L=W=100 are shown in Fig. 3.
\begin{figure}
\centerline {\epsfxsize=3.in\epsfbox{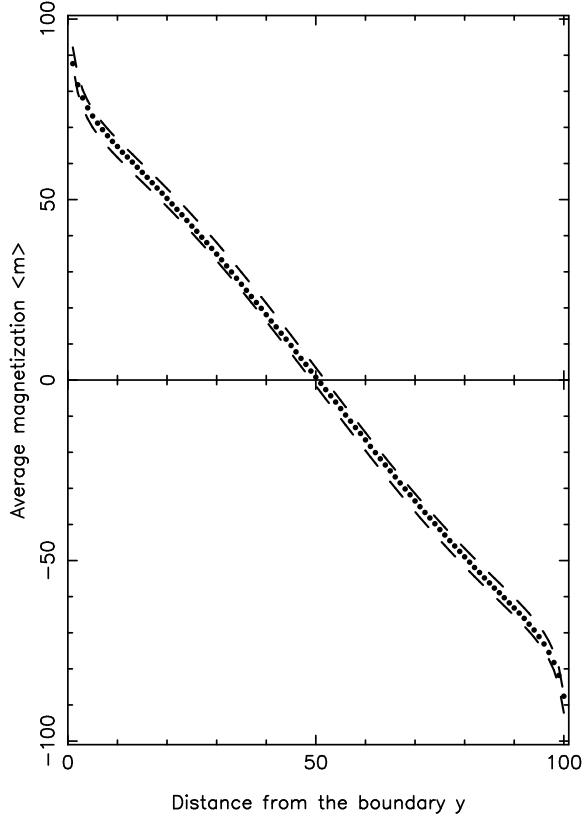}}
\vspace{3mm}
\caption{Average value of magnetization as a function
of the distance from the positive boundary for L=W=100.}
\end{figure}
The spins on the boundary y=0 were kept at the value of  +1, 
while the spins on the boundary
y=W were fixed at the value of -1.
The dashed lines are the  theoretical curve:
\bea
<m(r)> = C(\sin \pi\frac{y}{W})^{-\frac{1}{8}}\cos \pi\frac{y}{W},
\label{Eq:eq26}
\eea
obtained from Eq. (22), displaced as before.

*Percolation on infinite periodic strip. 
As in the Ising model, we approximate the infinite strip with 
a square  (0$\le$ x$\le$ 100,0$\le$ y $\le$ 100) which is periodic in x direction.
The results are shown in Fig. 4.  
\begin{figure}
\centerline {\epsfxsize=\columnwidth\epsfbox{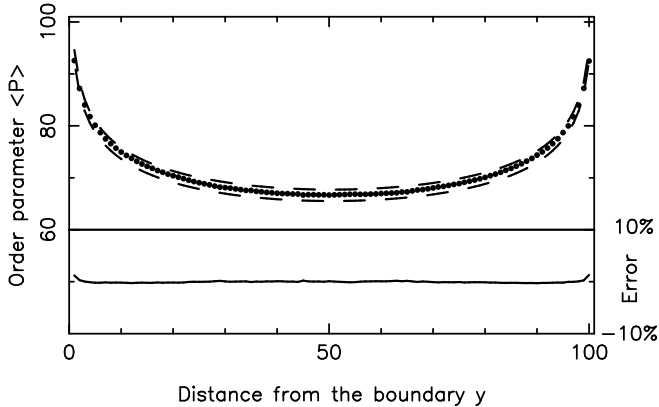}}
\vspace{3mm}
\caption{Order parameter as a function
of the distance from the boundary for a periodic square 100x100.}
\end{figure}
The dashed lines are the  theoretical curve:
\bea
<m(r)> = C(\sin \pi\frac{y}{W})^{-\frac{5}{48}}.
\label{Eq:eq27}
\eea

*Ising model on rectangle. 
The rectangle 
(0$\le$ x$\le$ L,0$\le$ y $\le$ W) was defined by spins fixed at the 
value of +1
on all boundaries. The average magnetization was measured as a function of
the distance from the boundary y = 0.The results for the square (L,W)=(100,100)
are shown in Fig. 5 and in Fig. 6. The magnetization in Fig. 5 was
measured along the line ($x=10$,y) while in Fig. 6 the measurement
was along the line ($x=50$,y).
\begin{figure}
\centerline {\epsfxsize=\columnwidth\epsfbox{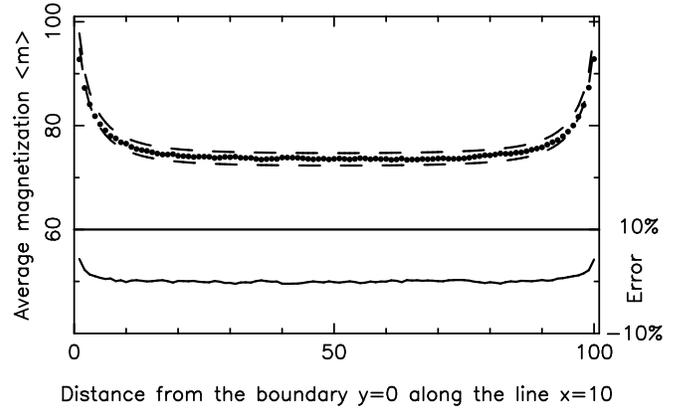}}
\vspace{3mm}
\caption{Average value of magnetization along the line (x=10,y)}
\end{figure}
\begin{figure}
\centerline {\epsfxsize=\columnwidth\epsfbox{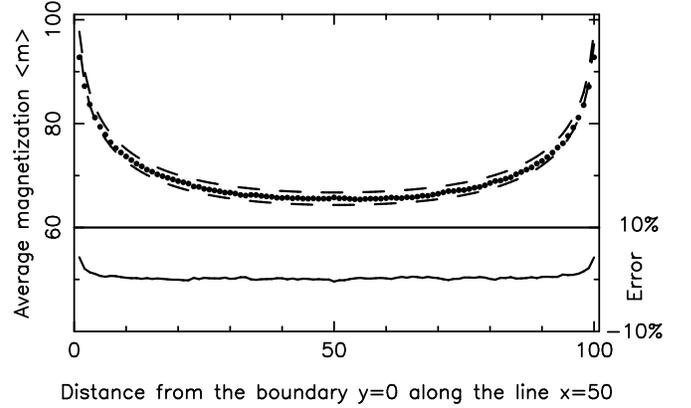}}
\vspace{3mm}
\caption{Average value of magnetization along the line (x=50,y)}
\end{figure}
The dashed lines are the  theoretical curve:
 \bea
<m(w)> = C[{{Im[sn({{2 K} \over {L}} w)]} \over {|cn({{2 K} \over {L}} w)dn({{2 K} \over {L}} w)|}}]^{-\frac{1}{8}},
\label{Eq:eq28}
\eea
obtained from Eq. (19).

*Percolation on rectangle. 
The rectangle is (0$\le$ x$\le$ 100,0$\le$ y $\le$ 100). 
The order parameter is given as a function of
the distance from the boudary y = 0.
The results along the line ($x=10$,y) are given 
in Fig. 7  while in Fig. 8 the measurement
was along the line ($x=50$,y).
\begin{figure}
\centerline {\epsfxsize=\columnwidth\epsfbox{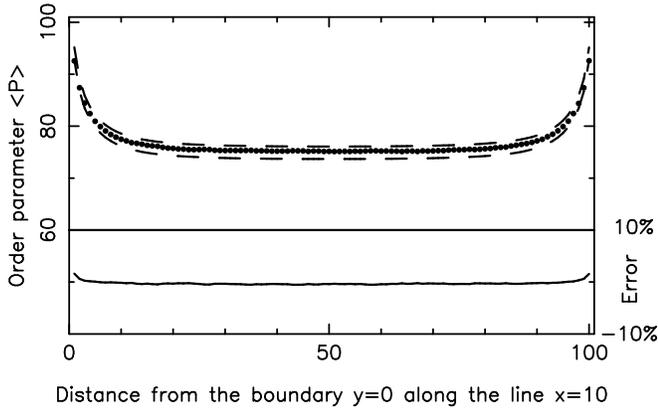}}
\vspace{3mm}
\caption{Order parameter along the line (x=10,y)}
\end{figure}
\begin{figure}
\centerline {\epsfxsize=\columnwidth\epsfbox{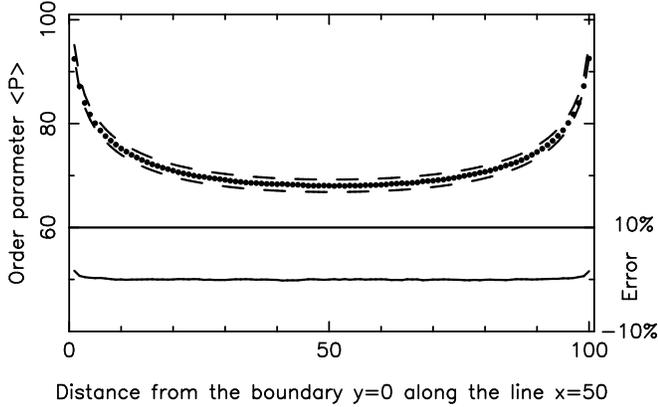}}
\vspace{3mm}
\caption{Order parameter along the line (x=50,y)}
\end{figure}
The dashed lines are the  theoretical curve:
 \bea
<m(w)> = C[{{Im[sn({{2 K} \over {L}} w)]} \over {|cn({{2 K} \over {L}} w)dn({{2 K} \over {L}} w)|}}]^{-\frac{5}{48}}.
\label{Eq:eq29}
\eea

*Ising model on the periodic rectangle. 
We used the rectangle 
(0$\le$ x$\le$ L,0$\le$ y $\le$ W) periodic in the x direction, 
with the spins kept at the value of +1 on the boundaries 
y=0 and y=W. As we have seen above, 
the case L$\ge$W resembles the infinite periodic strip. Here we will consider 
L$<$W. In Fig. 9 we show the average value of magnetization for 
rectangle L=80,W=101. In Fig. 10 the rectangle is L=40,W=101.
\begin{figure}
\centerline {\epsfxsize=\columnwidth\epsfbox{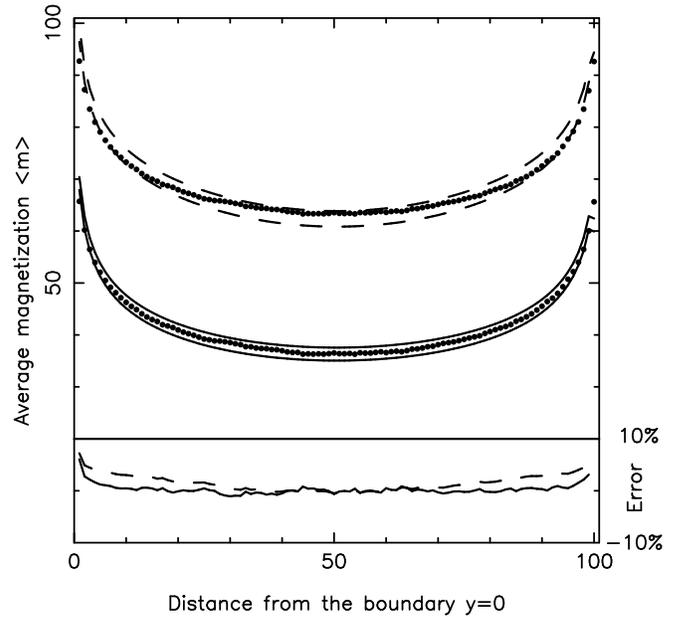}}
\vspace{3mm}
\caption{Average value of magnetization for the (80,101) rectangle.
The data is displayed twice, with a vertical displacement.
The upper set of lines compares the magnetization to the
prediction (23) based on the Gaussian model.  The lower set
is Eq. (24), supplied by conformal field theory.}
\end{figure}
\begin{figure}
\centerline {\epsfxsize=\columnwidth\epsfbox{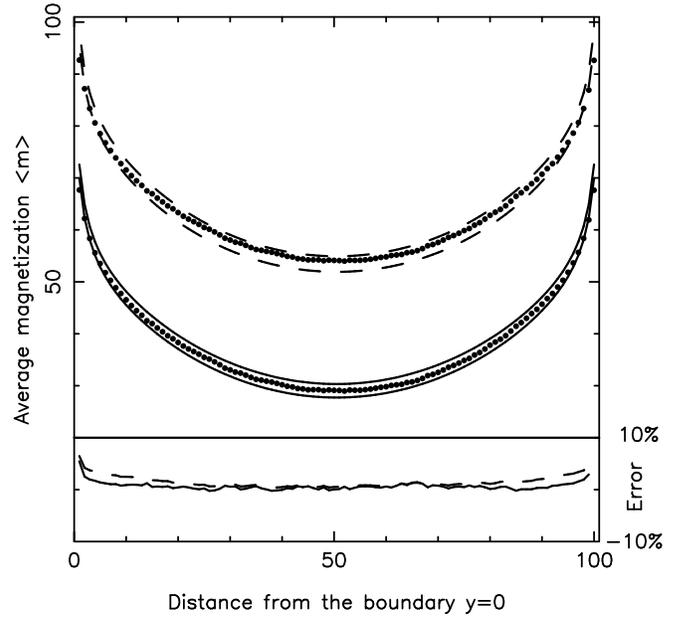}}
\vspace{3mm}
\caption{Average value of magnetization for the (40,101) rectangle.
Same comparisons as in figure 9.}
\end{figure}
In the lower part of the figure, the full line is the relative error  
for the conformal field theory prediction while the Gaussian model
error is represented by the dashed line. The same 
convention will be used in Fig. 11 and in Fig. 12.  
The agreement between the data and the predictions of the
Gaussian model is not very good; Eq. (24) coming from 
conformal field theory does much better.

*Percolation on the periodic rectangle. 
The position of percolation theory in conformal field theory is as yet
unclear: it apparently is not a unitary minimal model, and it is 
not known whether there is a differential equation to which $m(\vec r)$
is a solution.  In view of the failure of the Gaussian model
to describe the magnetization profile of the Ising model,
we have no theory for this case.  We have chosen to give the
same analysis as was done for the Ising model, changing only
the exponent $\Delta$.  Thus the top set of curves 
in Figures 11 and 12 compare the
behavior of the percolation order parameter to the Gaussian
model (23), while the lower set compare the same data to 
\bea
<m(x,y)>=
({{|\theta_{2} ({{\pi i} \over {L}} y )|^{\frac {1}{2}}
+|\theta_{3} ({{\pi i} \over {L}}} y )|^{\frac {1}{2}}
\over
{|\theta_{1} ({{2 \pi i} \over {L}} y)|^{  \frac {1}{8}}}})^{\frac {5}{6}}, 
\label{Eq:eq30}
\eea

which is related to the Ising model expression (24) through
the universality hypothesis Eq. (5).  There's no reason
for this to work!  But it seems to agree with the data!

\begin{figure}
\centerline {\epsfxsize=\columnwidth\epsfbox{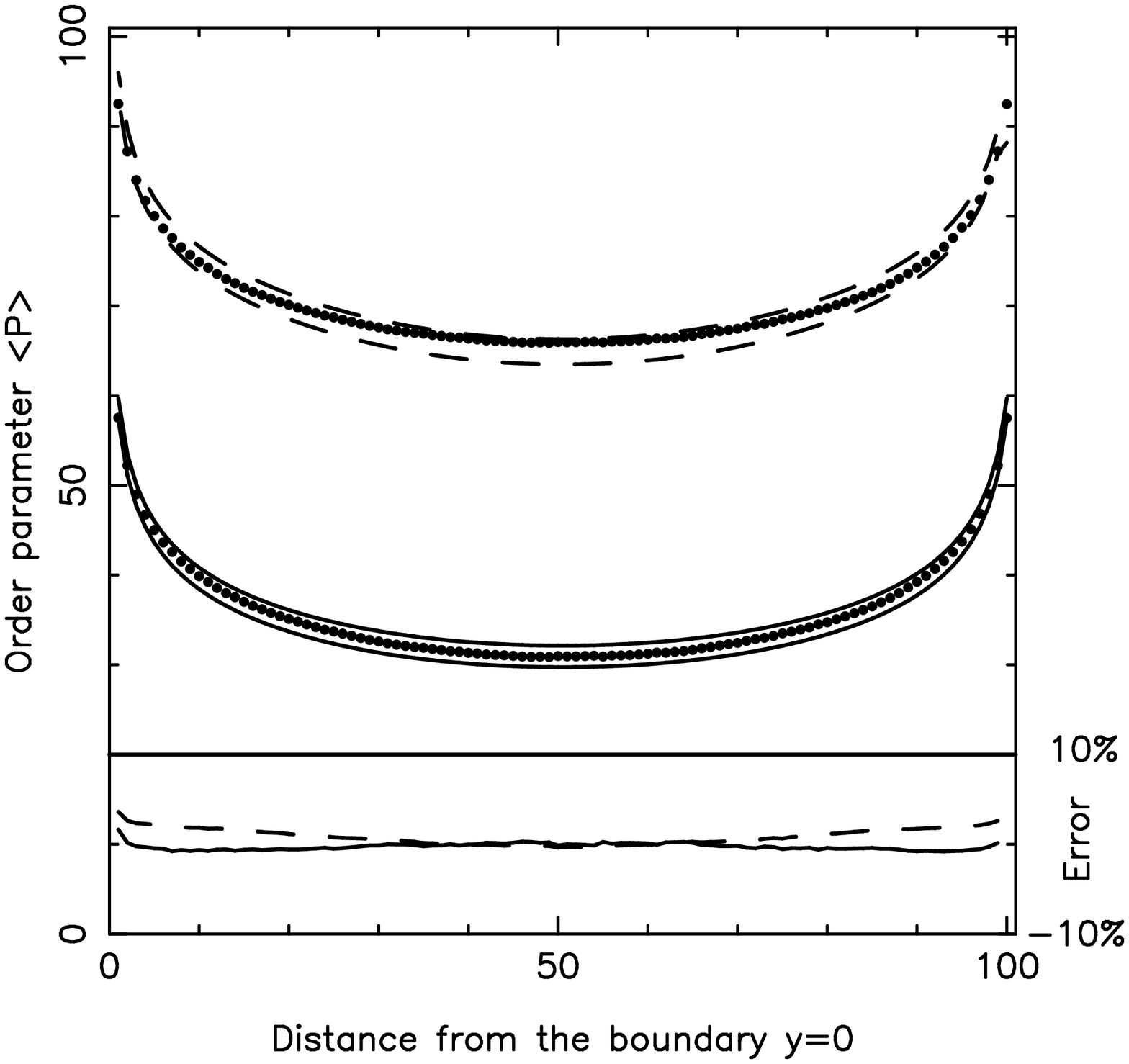}}
\vspace{3mm}
\caption{Order parameter for the (81,101) rectangle}
\end{figure}
\begin{figure}
\centerline {\epsfxsize=\columnwidth\epsfbox{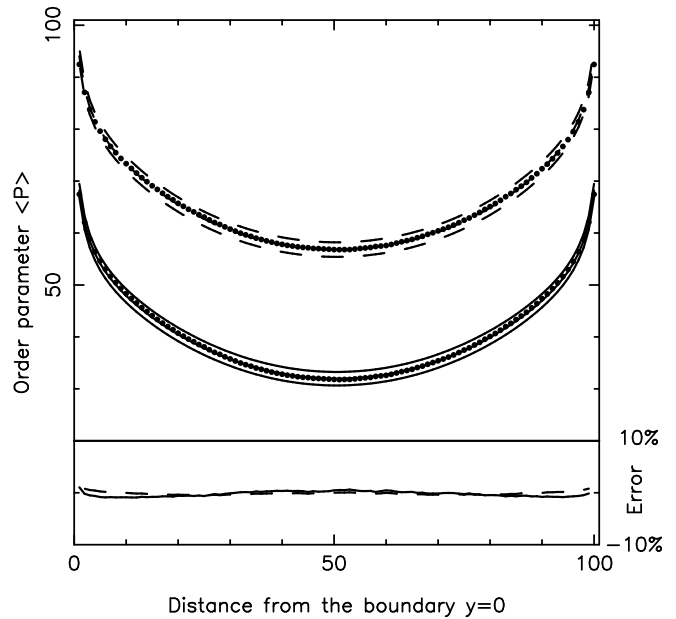}}
\vspace{3mm}
\caption{Order parameter for the (41,101) rectangle}
\end{figure}

\section{Planar spin model}

The planar spin model consists of unit spins that can point in any 
direction in the x-y plane. The Hamiltonian for this model can be written as

\bea
H=-J\sum_{<i,j>} cos(\theta_i-\theta_j ),
\label{Eq:eq31}
\eea
where the sum is over the nearest neighbors.
$\theta_i$ is the angle the {\em i} th spin makes with the x axis.

The model undergoes a Kosterlitz-Thouless (KT) transition \cite{kost,ko,tobche} at 
the critical temperature $T=T_c$. 
Below this temperature there is algebraic order at all
temperatures, implying a lack of a length scale
and thus conformal invariance; unlike three dimensional
phase transitions, however, the
exponent $\eta$ 
depends on temperature.
We determined $m ( {\vec r} )$ by means of simulations
on a disk of radius 100, with $\theta = 0$ on the boundary 
(i.e. the spins on the boundary are constrained to point
in the x direction), using Wolff's 
algorithm\cite{wolff}.
After the equilibration we performed 10$^6$ Monte Carlo steps per site.
By using $<m(r)>$ obtained from numerical simulations  we can find $\eta$,
since
Eqs.(3) and (17) imply
\bea
\ln<m(r)> = -\frac{\eta}{2}ln(A-\frac{r^{2}}{A})+constant.
\label{Eq:eq32}
\eea
 
Linear regression was used 
to determine the slopes of the lines in Fig. 13, 
which gives the exponent $\eta (T).$

\begin{figure}
\centerline {\epsfxsize=3.in\epsfbox{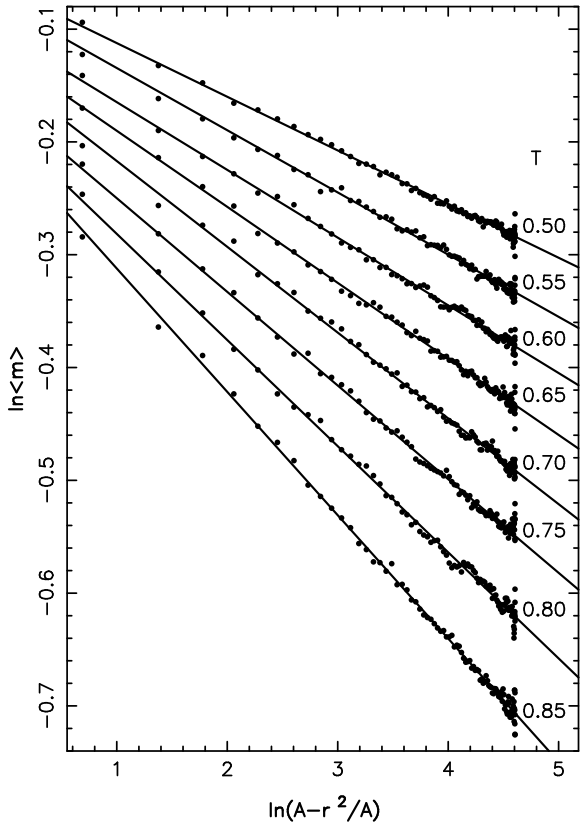}}
\vspace{3mm}
\caption{Linear fits for temperatures 0.85J, 0.80J, 0.75J, 
0.70J, 0.65J, 0.60J, 0.55J, 0.50J}
\end{figure}

Our results are given in Table \ref{table}:

\begin{table}[h]
\begin{center}
\begin{tabular}{|c|cccccccc|} 
T & 0.85J & 0.80J & 0.75J & 0.70J & 0.65J & 0.60J  
& 0.55J & 0.50J \\  \hline
$\eta$ & 0.2187 & 0.1882 & 0.1660  &  0.1521 
& 0.1348 & 0.1201 & 0.1103 & 0.0951     \\ 
\end{tabular}
\caption{\label{table}Critical exponent as a function of temperature}
\end{center}

According to Kosterlitz-Thouless theory\cite{chaik}, the long
wavelength spin fluctuations are described by
the Hamiltonian (32) with a renormalized coupling
constant J(T), given by
\bea
J(T)/T = {\frac 2 \pi} - (constant) \sqrt{{T-T_C} \over {T_C}},
\label{Eq:eq33}
\eea
where the constant is determined by the core energy
for a vortex.
This determines the temperature dependence of $\eta = T/2 \pi J$,
so that close to the Kosterlitz-Thouless transition

\bea
[{\frac{1}{4}}-\eta (T)]^2\approx \frac{T_C-T}{T_C},
\label{Eq:eq34} 
\eea
implying that $\eta$ approaches its critical value by a
square-root cusp.  We note that many of the previous 
studies\cite{tobche,FFS} of
$\eta (T)$ have incorrectly assumed a linear
dependence.

Figure 14 shows the analysis of the data from Table
\ref{table}.
\end{table}

\begin{figure}
\centerline {\epsfxsize=3.in\epsfbox{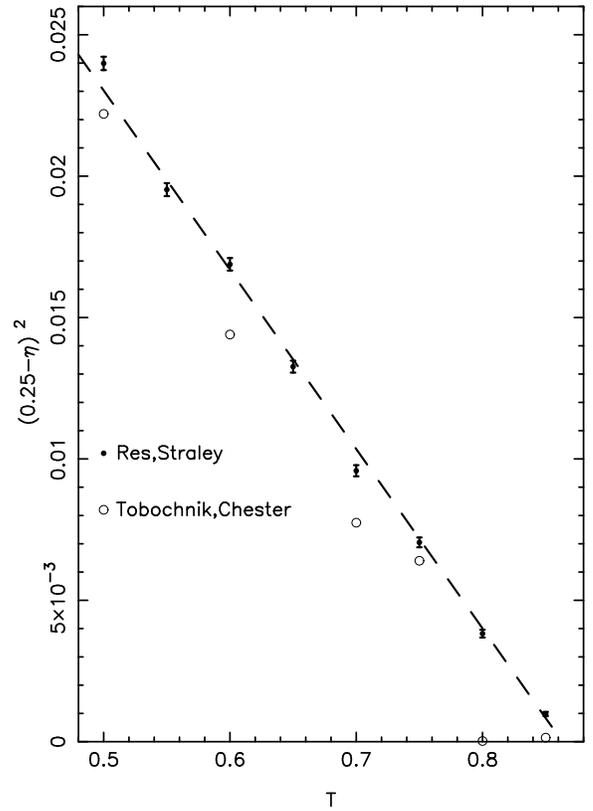}}
\vspace{3mm}
\caption{Data comparison}
\end{figure}

The points follow a straight line as predicted by eq.(34). The critical temperature is
given by the intersection of this line with the temperature axis and is found to be
\bea
T_C / J = 0.863 \pm 0.009.
\label{Eq:eq35} 
\eea
In Fig. 14  we also plotted the results obtained by Tobochnik and Chester (TC) \cite{tobche}.
We notice that their data consistently fall below our line, but are 
consistent with it.
TC analyzed their data using a linear extrapolation to obtain
$T_{C} /J=0.89 \pm 0.01$.
Our reanalysis of their data indicates a lower value.
We also quote the estimates for $T_C$ obtained by some other authors:
Weber and Minnhagen \cite{min} give $T_C / J = 0.887 \pm 0.002$
while Olsson \cite{ols} finds $T_C / J = 0.8922(2)$.  

The results obtained by computer simulations in the critical region 
are affected by the finite size effects - one is  trying to obtain predictions for the infinite system
by simulating a finite system.
Our approach avoids this problem because the theory says that the critical exponent $\eta$
for the model defined on a disk must in principle be equal to the exponent $\eta$
for the infinite system. The only approximation involved is the passage 
from a discrete system to continuum.
\section{Summary}

 In this work, we studied the magnetization of the Ising model
and percolation defined on finite geometries. We derived the
order parameter for rectangle by using conformal symmetry.
We showed how the results previously obtained by using 
conformal field theory can be obtained by studying a simple 
Gaussian model. We numerically studied order parameter in various
geometries by using Monte Carlo methods and compared analytical 
results with the outcome of our simulations.

\section{Appendix A}
To determine the order parameter for the model defined on an infinite 
strip ( x, 0$\le $y$\le$W) with spins fixed at the value of +1 on
 the boundaries,
we find the electrostatics Green function for the infinite strip with
Dirichlet boundary conditions:
\bea
G( Z , z ) = {\frac {1} {2 \pi }} (\ln | \sin {{\pi | Z - z^{*} |
} \over {2W}} |)
 - \ln | \sin {{\pi | Z - z |} \over {2W}} |),
\label{Eq:eq36}
\eea
where $z=x+iy$ is the complex representation of the plane, and
$z^{*} = x-iy$ is the position of an image of $z$\cite{jack}.
This result can be obtained by conformal mapping from the half-plane; or
from the observation that Eq.(36) has logarithmic singularities of
opposite signs at the positions of all images (in the two boundaries) of
$z$ and is a solution to the two-dimensional Laplace problem elsewhere,
since it can be written as the sum of a function of $Z$ and a function of
$Z^{*}$.
The non-singular part of the Green function is then
\bea
\tilde G(\vec r)=\frac {1}{2\pi}ln(\sin \frac {\pi y}{W}).
\label{Eq:eq37}
\eea
Eq. (15) then gives the result quoted in text.
 
The magnetization profile for a rectangle can be found using a
conformal mapping, as follows:
The semi-infinite complex plane $z=(x,y \ge 0)$ is mapped onto the rectangle 
$(- {\frac{L}{2}} \le u \le {\frac{L}{2}},0 \le iv \le iW )$
in the complex plane $w=(u,iv)$
by the Schwarz-Christoffel transformation\cite{chur,defra,burr}
\bea
w(z)=- {{L} \over {2K}}\int_{0}^zg(z^\prime)dz^\prime,
\label{Eq:eq38}
\eea
where
\bea
g (z)=\frac {1}{\sqrt{1-k^2 z^2}}\frac {1}{\sqrt{1- z^2}},
\label{Eq:eq39}
\eea
with
\bea
K=\int_{0}^1|g(x^\prime)|dx^\prime,
\label{Eq:eq40}
\eea
\bea
K^\prime=\int_{1}^a|g(x^\prime)|dx^\prime.
\label{Eq:eq41}
\eea
The ratio of $K$ and $K^\prime$  is related to the nome 
q of the Jacobian elliptic functions
\cite{abr}
\bea
q=e^{-\pi \frac {2W}{L}}.
\label{Eq:eq42}
\eea
The inverse of the function $w(z)$ is the Jacobian elliptic function 
$z=sn({{2K}\over{L}} w)$,
which implicitly depends \cite{abr} on the nome q.
Then we can put (4) and (38) into Eq. (1), noting that
$y=Im[sn({{2K}\over{L}} w)]$,
to obtain
 \bea
<m( w)> = C|\frac{Im[sn({{2K}\over{L}} w)]}{|\sqrt{1-k^2 sn^2({{2K}\over{L}} w)}\sqrt{1-sn^2({{2K}\over{L}} w)}|}|^{-\Delta}.
\label{Eq:eq43}
\eea
We use the relations between Jacobi elliptic functions sn, cn, dn\cite{abr}
to obtain Eq. (19).

This result can also be obtained through the connection
with the Gaussian model.  This requires finding the 
Dirichlet Green function for the rectangle, which 
can be found by the method of images.  
The images of the point $\vec r$ lie on four rectangular
Bravais lattices, and the Green function itself is
an infinite sum of the corresponding logarithmic terms.
The sum can be evaluated by observing
that $exp(2 \pi G(\vec R,\vec r))$ has only simple poles and zeroes, which form
periodic arrays, and thus it is a function of elliptic type\cite{abr}.  Inspection 
then gives
\bea
exp(2 \pi G(\vec R, \vec r)) = {{\theta_{1}(\pi {{Z-z} \over {2L}})
\theta _{1}(\pi{{Z-z^{\prime \prime}}\over {2L}})}
\over {\theta_{1}(\pi {{Z-z^{\prime}}\over {2L}}) \theta_{1}(\pi {{Z-z^{\prime \prime \prime}}\over {2L}}) }},
\label{Eq:eq44}
\eea
where $\theta_{1}$ is the theta function\cite{modern,abr}, $Z$ is the complex
representation of $\vec R$, and the $z$'s are the complex representations 
of $\vec r$ and the image points  
$z^\prime = z^*$,
$z^{\prime \prime}= L-z$,
$z^{\prime \prime \prime}= L-z^*$.
The resulting 
expression for the order parameter profile is
\bea
<m(x,y)>=|\frac {\theta_{2}(\frac {\pi x}{L})
\theta_{1}(\frac {i\pi y}{L})}{\theta_{2}(\pi \frac {x+iy}{L})}|^{-\Delta}.
\label{Eq:eq45}
\eea
This representation is equivalent to the one used above,
but is easier to evaluate numerically, since the series 
representations
\cite{abr}
for 
the theta functions converge extremely rapidly.

The Dirichlet Green function for a $L \times W$ rectangle  
with fixed boundary conditions on two sides and periodically
connected along the other two
has logarithmic singularities at $Z = z + m L + i n W$, and
logarithmic singularities of opposite sign at
$Z = z^{*} + mL + inW$, where $z^{*} = x - iy$.  This
behavior is reproduced by the
function
\bea
G(Z,z)=-\frac{1}{2\pi}ln|\frac{\theta_{1}(\pi {{Z-z}\over {L}})e^{\frac{\pi}{L^{2} \tau}(Z-z)^2}}
{\theta_{1}(\pi {{Z-z^*}\over{L}})e^{\frac{\pi}{L^{2}\tau}(Z-z^*)^2}}|.
\label{Eq:eq46}
\eea
Here $\tau = 2W/L$ specifies the shape of the rectangle.
We take the limit $Z\mapsto z$ as we did in the case of 
nonperiodic rectangle and obtain eq. (23).

\section{Appendix B}

According to conformal field theory\cite{defra}, the N-point
correlation functions of many critical models can be found by
solving a differential equation originating in the algebra
of the operators corresponding to the fields.  
In many cases, the correlation function factors into a product
of holomorphic (depending on $z$) and antiholomorphic ($z^{*}$)
parts, or a sum of a small number of such terms.
Cardy\cite{cardy2} has given a prescription for introducing
a straight homogeneous boundary into this formalism.  In
effect, the N-point function is found by solving the 
2N-point differential equation for the holomorphic part,
and then evaluating this with the second set of points
at the image position (i.e. $z_{N+1} = z_{1}^{*}$ if the
boundary is the real axis).  Since the differential equations
are of higher than first order, there are several solutions
to choose from, and the choice is determined by the boundary
condition (e.g.Dirichlet) desired.

The problem we wish to consider is the rectangle with Dirichlet
boundary conditions on two opposite edges and periodic continuation
on the other two edges.  This is equivalent to a finite length 
cylinder with Dirichlet boundary conditions on the ends, or to
a torus that is cut by a single Dirichlet seam.

The differential equation appropriate for the Ising model spin
operator on a torus has been given by Eguchi and Ooguri\cite{eguchi},
and the solutions for the two-point function are given by Di Francesco et al.\cite{DSZ}.
They all have the form
\bea
<\sigma \sigma> =  {{[\theta_{\nu}({{z_{1}-z_{2}}\over {2}})]^{\frac {1}{2}}}
\over {[\theta_{1}(z_{1}-z_{2})]^{\frac {1}{8}}}},
\label{Eq:eq47}
\eea
where $\theta_{\nu}$ ($\nu = 2,3,4$) are the theta functions with the
periodicity of the torus.
The combination $<\sigma \sigma>_{2} + <\sigma \sigma>_{3}$
has the desired periodicity in Im z.
We then follow Cardy by evaluating this
with $z_{1} = z$, $z_{2} = z^{*}$.


\begin{references}

\bibitem{defra} P. Di Francesco, P. Mathieu, D. S\'{e}n\'{e}chal, 
 {\it Conformal Field Theory}, (Springer-Verlag, 1997).

\bibitem{cardy} J.Cardy, {\it Scaling and Renormalization in Statistical 
Physics}, (Cambridge University Press, 1996).

\bibitem{tob} H. Gould, J. Tobochnik,
 {\it An Introduction to Computer Simulation Methods}, 
(Addison-Wesley, 1996).

\bibitem{burr} T. W. Burkhardt, E.Eisenriegler, J. Phys. A {\bf 18}, L83(1985).
\bibitem{3D} Higher-dimensional systems can also be conformally invariant
at the critical point.  The set of conformal transformations is
very small, so that there are few consequences; but we note that
Eq. (4) continues to describe the decay of correlations near an
ordered boundary, and that there is a conformal transformation of
the half-space into the sphere interior, so that Eq. (17) is also
valid.  This still isn't very interesting, because the general
case of Eq. (3) is $\Delta = \frac {1}{2} (d-2 + 
\eta)$, where $\eta$ is expected to be small.

\bibitem{abr} M. Abramowitz, I. A. Stegun, {\it Handbook of Mathematical 
Functions}, (Dover, New York, 1964), Chapter 16.

\bibitem{burg} T. W. Burkhardt, I. Guim, Phys. Rev. B {\bf 53}, 2080 (1987). 

\bibitem{swang} R. H. Swendsen, J.-S. Wang, 
Phys. Rev. Lett. {\bf 58}, 86 (1987).

\bibitem{chur} R. V. Churchill, J.W. Brown, 
{\it Complex Variables and Applications}, (McGraw-Hill, 1990).

\bibitem{stauf} D. Stauffer, A. Aharony,
{\it Introduction to Percolation Theory }, (Taylor \& Francis, 1994). 

\bibitem{modern} E. T. Whittaker, G. N. Watson, {\it A Course of Modern Analysis},
(Cambridge University Press, 1962) 

\bibitem{jack} J. D. Jackson, {\it Classical Electrodynamics}, 
(Willey, 1975)

\bibitem{kost} J. M. Kosterlitz, D. J. Thouless, 
J. Phys. C {\bf 6}, 1181(1973).

\bibitem{ko} J. M. Kosterlitz, J. Phys. C {\bf 7}, 1046(1974).

\bibitem{tobche} J. Tobochnik, G. W. Chester, Phys. Rev. B {\bf 20}, 3761(1979).

\bibitem{wolff} U. Wolff, Phys. Rev. Lett. {\bf 62} 361(1989).


\bibitem{chaik} P. M. Chaikin, T. C. Lubensky{\it Principles of Condensed
Matter Physics}, (Cambridge University Press, 1995).
 
\bibitem{FFS} J. F. Fern\'andez, M. F. Ferreira, and J. Stankiewicz,
Phys. Rev. B {\bf 34}, 292 (1986).
\bibitem{min} H. Weber, P.Minnhagen, Phys. Rev. B {\bf 37}, 5986 (1988). 
\bibitem{cardy2}
J. L. Cardy, Nucl. Phys. {\bf B240} [{\bf FS12}] 514 (1984).
\bibitem{eguchi}T. Eguchi and H. Ooguri, Nucl. Phys. {\bf B282},
308 (1987).
\bibitem{DSZ} P. Di Francesco, H. Saleur, and J.-B. Zuber, Nucl. Phys.
{\bf B290} [{\bf FS20}], 527 (1987).
\bibitem{ols} P. Olsson, Phys. Rev. Lett. {\bf 73}, 3339 (1994).
\end{references}
\end{document}